\documentclass[aps,prl,a4paper,nofootinbib,showpacs,showkeys,
preprintnumbers,superscriptaddress,twocolumn]{revtex4}

\usepackage{amsmath}
\usepackage{amssymb}
\usepackage{bm}
\usepackage{graphicx}
\usepackage{color}

\begin{document}

\title{Elliptic Flow from Non-equilibrium Initial Condition with a Saturation Scale}

\author{M. Ruggieri}\email{marco.ruggieri@lns.infn.it}
\affiliation{Department of Physics and Astronomy, University of Catania, Via S. Sofia 64, I-95125 Catania}

\author{F. Scardina}
\affiliation{Department of Physics and Astronomy, University of Catania, Via S. Sofia 64, I-95125 Catania}
\affiliation{INFN-Laboratori Nazionali del Sud, Via S. Sofia 62, I-95123 Catania, Italy}

\author{S. Plumari}
\affiliation{Department of Physics and Astronomy, University of Catania, Via S. Sofia 64, I-95125 Catania}
\affiliation{INFN-Laboratori Nazionali del Sud, Via S. Sofia 62, I-95123 Catania, Italy}

\author{V. Greco}\email{greco@lns.infn.it}
\affiliation{Department of Physics and Astronomy, University of Catania, Via S. Sofia 64, I-95125 Catania}
\affiliation{INFN-Laboratori Nazionali del Sud, Via S. Sofia 62, I-95123 Catania, Italy}


\begin{abstract}
A current goal of relativistic heavy ion collisions experiments
is the search for a Color Glass Condensate (CGC) as the limiting state of QCD matter at very high density.
In viscous hydrodynamics simulations, a standard Glauber initial condition leads to estimate $4\pi \eta/s \sim 1$, while
employing the Kharzeev-Levin-Nardi (KLN) modeling of the glasma leads to at least a factor of 2 larger $\eta/s$.
Within a kinetic theory approach based on a relativistic Boltzmann-like transport simulation, 
our main result is that the out-of-equilibrium initial distribution 
reduces the efficiency in building-up the elliptic flow.
At RHIC energy we find the available data on $v_2$
are in agreement with a $4\pi \eta/s \sim 1$ also for KLN initial conditions.
More generally, our study shows that the initial non-equilibrium in p-space 
can have a significant impact on the build-up of anisotropic flow.

\end{abstract}


\maketitle

Ultra-relativistic heavy-ion collisions (uRHICs) at the Relativistic Heavy-Ion Collider (RHIC) and the Large Hadron Collider (LHC) 
create a hot and dense system of strongly interacting matter. In the last decade it has been reached a 
general consensus that such a state of matter is not of hadronic nature and there are several signatures
that it is a strongly interacting quark-gluon plasma (QGP) \cite{STAR_PHENIX, ALICE_2011, Science_Muller}. 
A main discovery has been that the QGP has a very small shear viscosity to density entropy, $\eta/s$,
which is more than one order of magnitude smaller than the one of water~\cite{Csernai:2006zz,Lacey:2006bc}, 
and close to the lower bound of $1/4\pi$ conjectured for systems at infinite strong coupling \cite{Kovtun:2004de}. 
A key observable to reach such a conclusion is the so-called elliptic flow,
$v_2 = \langle cos(2 \varphi_p) \rangle= \langle (p_x^2-p_y^2)/(p_x^2+p_y^2) \rangle$,
with $\varphi_p$ being the azimuthal angle
in the transverse plane and the average meant over the particle distribution. 
In fact, the expansion of the created matter
generates a large anisotropy of the emitted particles that can be primarily measured by $v_2$.  
Its origin is the initial spatial eccentricity, $\epsilon_x=
 \langle x^2-y^2\rangle/\langle x^2+y^2 \rangle$, of the overlap region in non-central collisions.
 The observed large $v_2$ 
 is considered a signal of a very small $\eta/s$
because it means that the system is very efficient in converting $\epsilon_x$
into an anisotropy in the momentum space $v_2$, a mechanism that would be strongly damped in a  
system highly viscous that dissipates and smooths anisotropies
\cite{Romatschke:2007mq, Heinz,Cifarelli:2012zz}.
Quantitatively both viscous hydrodynamics \cite{Romatschke:2007mq,Luzum:2008cw,Heinz,Song:2011hk,Schenke:2010nt,Niemi:2011ix}, 
and  transport Boltzmann-like approaches~\cite{Ferini:2008he, Xu:2008av,Xu:2007jv,Bratkovskaya:2011wp,Plumari_BARI}
agree in indicating an average $\eta/s$ of the QGP lying in the range $4\pi\eta/s \sim 1-3$.

The uRHIC program 
offers the tantalizing opportunity to explore the existence of an exotic state,
namely the Color Glass Condensate (CGC)~\cite{McLerran:1993ni,McLerran:1993ka},
see~\cite{Gelis:2010nm,Iancu:2003xm} for reviews.
Such a state of matter would be primarily generated by the very high density of the gluon parton distribution
function at low $x$ (parton momentum fraction), which triggers a saturation
of the gluon distribution function at a $p_T$ below the saturation scale, $Q_s$~\cite{Kovchegov:1999yj}.
Even if at first sight surprisingly, the study of the shear viscosity $\eta/s$ of the QGP 
and the search for the CGC are related.
In fact, the main source of uncertainty for $\eta/s$ comes from the unknown initial conditions 
of the created matter ~\cite{Luzum:2008cw} and confirmed later by further works~\cite{Alver:2010dn,Song:2011hk,Adare:2011tg}.  

A simple geometrical description through the Glauber model~\cite{Miller:2007ri} predicts a 
$\epsilon_x$ smaller at least 25-30$\%$ than the eccentricity of the CGC, for most of the centralities of 
the collisions, see for example results within the
Kharzeev-Levin-Nardi (KLN) model~\cite{Kharzeev:2004if,Hirano:2005xf,Drescher:2006pi}, 
factorized KLN (fKLN) model~\cite{Drescher:2006ca}, 
Monte Carlo KLN (MC-KLN) model ~\cite{Drescher:2006ca,Hirano:2009ah} and dipole 
model~\cite{Drescher:2006pi,Albacete:2010pg}.  
The uncertainty in the initial condition 
translates into an uncertainty on $\eta/s$ of at a least a factor of two as estimated
by mean of several viscous hydrodynamical approaches \cite{Luzum:2008cw,Alver:2010dn,Song:2011hk,Adare:2011tg}. 
More explicitely, the experimental
data of $v_2(p_T)$ at the highest RHIC energy are in agreement with a fluid at $4\pi\eta/s\sim 1$
according to viscous hydrodynamics simulation, assuming a standard Glauber
initial condition. Assuming an initial  fKLN or MC-KLN space distribution the comparison
favors a fluid at $4\pi\eta/s\sim 2$. The reason is the larger initial $\epsilon_x$ 
of the fKLN, which leads to larger $v_2$ unless a large $\eta/s$ is considered.
However, in~\cite{Alver:2010dn,Adare:2011tg} it has been shown that viscous hydrodynamics fails to 
reproduce both $v_2$ and $v_3=\langle cos(3 \varphi_p)\rangle$ if the same $4\pi\eta/s \sim 2$
is assumed. At variance a Glauber initial condition can account for both with the same $4\pi\eta/s=1$.
However the indirect effect on $\epsilon_x$ is not a unique
and solid prediction of the CGC modellings; for example, 
the approach based on the solution of the Classical Yang-Mills (CYM)
equations predict a somewhat smaller initial eccentricity~\cite{Lappi:2006xc,Schenke:2012wb}.
Very recently employing a $x-space$ distribution inspired to the CYM approach
in a viscous hydrodynamical approach \cite{Gale:2012rq} it has been show that not only $v_2$ but also higher harmonics 
can be correctly predicted with a  $4\pi\eta/s\sim 1.5$ instead of $\sim 2$, which is in qualitative
agreement with the fact that CYM tend to predict quite smaller $\epsilon_x$ with respect to fKLN.
However our present studies focus on the effect of the initial non-equilibrium in $p-$space,
an issue discarded in all previous studies including  the recent ones~\cite{Gale:2012rq,Schenke:2012wb}. 

In this Letter, we point out that the implementation of the melted CGC
in hydrodynamics takes into account only the different space distribution
with respect to a geometric Glauber model,
discarding the key and more peculiar feature of the damping
of the distribution for $p_T$ below the $Q_s$ saturation scale.
We have found by mean of kinetic theory
that this has a pivotal role on the build-up of $v_2$.

We adopt the model which was firstly introduced
by Kharzeev, Levin and Nardi~\cite{Kharzeev:2004if}
(KLN model) even if in the
regime of over saturation in $A+A$ collisions some aspects are better caught 
by a CYM approach~\cite{Blaizot:2010kh}.
In particular, to prepare the initial conditions of our simulations we refer to
the factorized-KLN (fKLN) approach as introduced in~\cite{Drescher:2006ca,Hirano:2009ah}.
This will allow for a direct comparison with viscous hydrodynamics results. 
in which the coordinate space distribution function of gluons arising from the melted CGC is assumed to be 
\begin{eqnarray}
\frac{d N_g}{dy d^2\bm{x}_\perp} = \int d^2 \bm{p}_T\, p_A(\bm x_\perp) p_B(\bm x_\perp)  \Phi({\bm p_T},{\bm x_\perp},y)~,
\label{eq:densityCS}
\end{eqnarray}
where $\Phi$ corresponds to the momentum space distribution in the $k_T$ factorization 
hypothesis~\cite{Gribov:1984tu,Kovchegov:2001sc},
\begin{eqnarray}
\Phi({\bm p_T},{\bm x_\perp},y) &=&\frac{4\pi^2 N_c}{N_c^2-1} \frac{1}{p_T^2}
\int^{p_T} d^2\bm{k}_T
\alpha_S(Q^2)\nonumber\\
&&\times\phi_A(x_1,k_T^2;\bm{x}_\perp)\nonumber\\
&&\times\phi_B(x_2,(\bm{p}_T -\bm{k}_T)^2;\bm{x}_\perp)~.
\label{eq:densityPS}
\end{eqnarray}
Here $x_{1,2} = p_T \exp(\pm y)/\sqrt{s}$ and 
the ultraviolet cutoff $p_T = 3$ GeV$/c$  assumed in the $p_T$ integral in Eq.~\eqref{eq:densityCS}; 
$\alpha_S$ denotes the strong coupling constant,
which is computed at the scale $Q^2 = \text{max}(\bm k_T^2,(\bm p_T - \bm k_T)^2)$ according to
the one-loop $\beta$ function but frozen at $\alpha_s=0.5$ in the infrared region
as in~\cite{ALbacete:2010ad,Hirano:2005xf,Albacete:2010pg}.
In Eq.~\eqref{eq:densityCS} $p_{A,B}$ denote the probability to find one nucleon
at a given transverse coordinate, $p_A({\bm x}_\perp) = 1-
\left(1-{\sigma_{in}} T_A(\bm x_\perp)/A\right)^A$,
where $\sigma_{in}$ is the inelastic cross section and  $T_A$ corresponds
to the usual thickness function of the Glauber model. 

The main ingredient to specify in Eq.~\eqref{eq:densityPS}  
is the unintegrated gluon distribution function (uGDF) for partons 
coming from nucleus $A$, which is assumed to be:
\begin{eqnarray}
\phi_A(x_1,k_T^2;\bm{x}_\perp) = \frac{\kappa \,Q_s^2}{\alpha_s(Q_s^2)}
\left[\frac{\theta(Q_s - k_T)}{Q_s^2 + \Lambda^2} + \frac{\theta(k_T - Q_s)}{k_T^2 + \Lambda^2}\right]~
\label{eq:phiA}
\end{eqnarray}
where we see the peculiar feature of the CGC that is the saturation of the distribution
for $p_T< Q_s$; a similar equation holds for partons belonging to nucleus $B$. 
Following~\cite{Drescher:2006ca} we take the saturation scale for the nucleus $A$ as
\begin{equation}
Q_{s,A}^2(x,{\bm x}_\perp)=2{\text{GeV}}^2\left(\frac{T_A({\bm x}_\perp)}{1.53 p_A({\bm x}_\perp)}\right)
\left(\frac{0.01}{x}\right)^\lambda~,
\label{eq:PPPppp}
\end{equation}
with $\lambda=0.28$, and similarly nucleus $B$. This choice 
is the one adopted in fKLN or MC-KLN and in hydro simulations~\cite{Drescher:2006ca,Song:2011hk}
to study the dependence of $v_2(p_T)$ on $\eta/s$.
Using Eqs.~\eqref{eq:PPPppp} and~\eqref{eq:densityCS} we find that $\langle Q_s\rangle \approx 1.4$ GeV
where the average is understood
in the transverse plane.

\begin{figure}
\begin{center}
\includegraphics[width=7.7 cm]{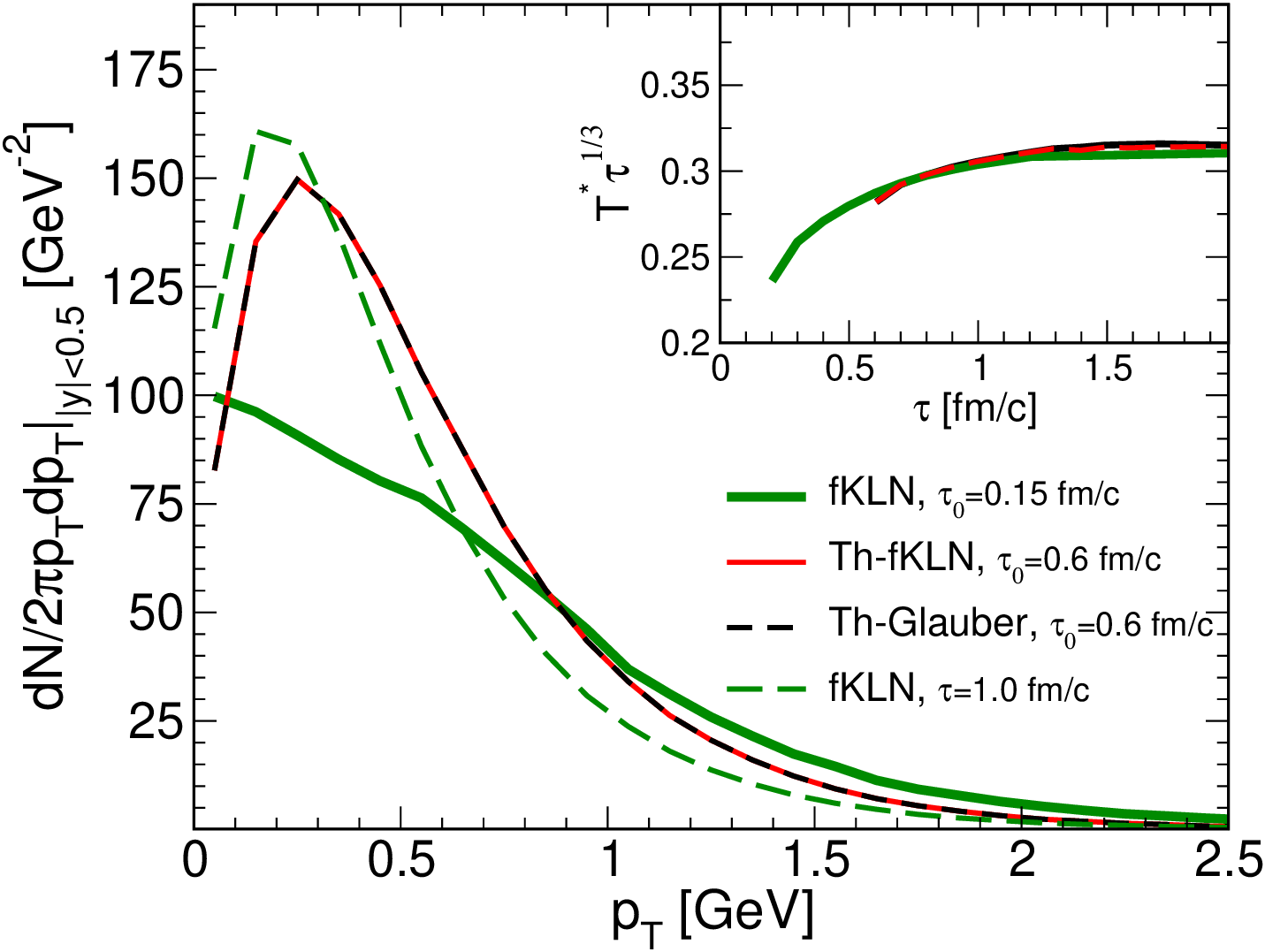}
\caption{\label{Fig:spettri_CGC}
Transverse momentum distributions within $|y|\leq 0.5$ 
for the case of fKLN model at the initial time (thick solid line)
compared to the the thermal one (thin solid and dashed lines) at their initial evolution time $\tau_0$
and to the one of fKLN after a dynamical evolution $\tau=1.0 \rm\, fm/c$ (green dashed line). 
{\em Inset}: time evolution of $\rm T \cdot \tau^{1/3}$ for the three different initial conditions considered
in the space rapidity $|\eta_s| \leq 0.5$.
All the panels refer to Au-Au collisions
at $\sqrt{s} = 200$ GeV, with an impact parameter $b=7.5$ fm.}
\end{center}
\end{figure}

We employ transport theory as a base of a simulation code of the fireball expansion created
in relativistic heavy-ion collision \cite{Ferini:2008he,Plumari:2010ah,Plumari_BARI,Plumari:2012xz}, 
therefore the time evolution of the gluons distribution function $f({\bm x}, {\bm p}, t)$
evolves according to the Boltzmann equation:
\begin{eqnarray}
p_\mu\partial^\mu f_1 &=& 
\int d\Gamma_2 d\Gamma_{1^\prime} d\Gamma_{2^\prime}
(f_{1^\prime}f_{2^\prime} - f_1 f_2)\nonumber\\ 
&&\times|{\cal M}|^2\delta^4(p_1 + p_2 - p_{1^\prime} - p_{2^\prime})~,
\label{eq:BE}
\end{eqnarray}
where $d^3 {\bm p}_{k} = 2 E_{k} (2\pi)^3 d\Gamma_k$ and ${\cal M}$ corresponds to the transition amplitude.

At variance with the standard use of transport theory, we have developed
an approach that, instead of focusing on specific microscopic calculations or modelings for the scattering matrix,
fixes the total cross section in order to have the wanted $\eta/s$. 
In Ref.\cite{Huovinen:2008te} it has been shown in 1+1D such an approach
is able to recover the Israel-Stewart viscous hydrodynamical evolution when $\eta/s$
is sufficiently small.
In 3+1D some of the authors has studied the analytical relation between $\eta$, temperature,
cross section and density and
as shown in \cite{Plumari:2012ep,Plumari:2012xz}, the Chapmann-Enskog approximation 
supplies such a relation with quite good approximation \cite{Wiranata:2012br}, in agreement with the results 
obtained using the Green Kubo formula. 
Therefore, we fix $\eta/s$ and compute the pertinent total cross section
by mean of the relation
\begin{equation}
\sigma_{tot}=\frac{1}{15} \frac{\langle p\rangle}{\rho \, g(a)} \frac{1}{\eta/s}~, 
\label{eq:sigma}
\end{equation}
which is valid for a generic differential cross section $d\sigma/dt \sim \alpha_s^2/(t-m_D^2)^2$
as proved in~\cite{Plumari:2012ep}.
In the above equation $a=T/m_D$, with $m_D$ the screening mass regulating the angular dependence
of the cross section, while  
\begin{eqnarray}
 g(a)=\frac{1}{50}\! \int\!\! dyy^6
\left[ (y^2{+}\frac{1}{3})K_3(2 y){-}yK_2(2y)\right]\!
h\left(\frac{a^2}{y^2}\right),
\label{g_CE}
\end{eqnarray}
with $K_n$ the Bessel function and $h$ corresponding to the ratio of the transport and the
total cross section. The maximum value of $g$, namely $g( m_D \rightarrow \infty)=h(m_D \rightarrow \infty)=2/3$, 
is reached for isotropic cross section; a smaller value of $g(a)$
means that a higher $\sigma_{tot}$ is needed to reproduce the same value of $\eta/s$. 
However, we notice that in the regime were viscous hydrodynamic applies 
(not too large $\eta/s$ and $p_T$)
the specific microscopic detail of the cross section is irrelevant and our approach 
is an effective way to employ transport theory to simulate a fluid at a given $\eta/s$.
From the operative point of view, keeping $\eta/s$ constant in our simulations
is achieved by evaluating locally in space and time the strength of the cross
section by means of Eq.~\eqref{eq:sigma}, where both parton densities and
temperature are computed locally in each cell. 
To realize a realistic freeze-out, when the local energy density reaches the cross-over region,
the $\eta/s$ increases linearly to match the estimated hadronic viscosity, 
as described in~\cite{Plumari_BARI,Plumari:2012xz}; 
this affects in the same way all the cases considered in the following.
 
In the following, we will consider three different types of initial distribution function in the phase-space,
two of which are the one employed till now for the investigation of the $\eta/s$, while the third one
is the genuine novelty of the present study. Furthermore, we refer to $Au+Au$ collision at $\sqrt{s}= 200 \, AGeV$
and $b=7.5 \rm \, fm$.
In this case, our result for initial eccentricity in the fKLN model is $\epsilon_x=0.357$
(which is in agreement with MC-KLN \cite{Drescher:2006ca} result used in hydro simulations). 
The standard initial condition for simulations of the plasma fireball created at RHIC
is a $\bm x$-space distribution given by the Glauber model 
and a $\bm p$-space thermalized spectrum in the transverse plane at a time $\tau = 0.6\rm \, fm/c$ with
a maximum initial temperature $\rm T_0 = 340 \rm MeV$. In this case, for
a standard mixture of $N_{part}$ and $N_{coll}$
we find $\epsilon_x=0.284$.
We will refer to this case as {\it Th-Glauber}.
Instead the study of the impact of an initial CGC state has been performed considering an $\bm x$-space
distribution given by the fKLN (or MC-KLN), while in the momentum space 
the spectrum has been considered thermalized at $\tau_0 \sim 0.6 \rm fm/c$; we refer to this case
as {\it Th-fKLN} and it is represents the case implemented in hydrodynamics, that has lead to the conclusion
that the CGC suggests a $4\pi \eta/s \sim 2$ \cite{Luzum:2008cw,Alver:2010dn,Song:2011hk,Adare:2011tg}. 
The third initial conditions is the full fKLN initial conditions where,
beyond the $\bm x$- space, the saturated distribution in $\bm p$-space
is implemented as well, see
Fig. \ref{Fig:spettri_CGC} solid thick line. As initial time we take $\tau_0=0.15 \rm fm/c$ 
because in this case 
there is no pre-assumption of thermalization.
This is not usually considered in hydrodynamics because there it is implicitly assumed a distribution function in $\bm p$-space
in local equilibrium, at least in the transverse plane. 
The choice of $\tau_0$ for the glasma-like initial condition is inspired by the recent results
of~\cite{Gelis:2013rba,Fukushima:2013dma,Ryblewski:2013eja} where it is discussed that, 
even if at $\tau=0^+$ the longitudinal pressure of the glasma
is negative, within a time $\tau\approx 1/Q_s$ it becomes positive thus making a description of the expanding system
in terms of a partonic distribution function quite reliable.
For all the previous cases, as usually done, a  Bjorken scaling at the initial time is assumed, identifying momentum rapidity $y_z$ 
with space rapidity $\eta_s=arctg^{-1}(z/\tau)$. 
For all the case considered the multiplicity $dN/d\eta$ at mid rapidity has been fixed initially equal to 400 
to approximately match the experimental data that for the impact parameter considered corresponds to about
$N_{part}=150$ in \cite{Alver:2010ck}.

In Fig.~\ref{Fig:spettri_CGC} we plot the initial spectra for the fKLN (thick solid line),
Th-fKLN (dashed line) and 
Th-Glauber (thin solid line) at their respective initial
times $\tau_0$, and the spectrum of the fKLN model after a time evolution $\Delta\tau=0.8 \rm \, fm/c$ (dashed green line). 
We notice that initially the fKLN spectrum is quite far from a thermalized spectrum; in fact,
it embeds the saturation effects which are proper of the melted CGC.
Neverthelsess, the spectrum evolves
to a thermalized one within $1$ fm/c.
Such a feature is confirmed by the inset of Fig. \ref{Fig:spettri_CGC}, 
where the quantity $T^* \cdot \tau^{1/3}$ is shown, with $T^*=E/3N$ representing,
the temperature in the case of a thermalized system. It is known that in the case of 1D expansion 
a thermalized system should keep such a ratio constant. 
We find that in the case of the fKLN (solid green line) the product $T^* \cdot \tau^{1/3}$ is strongly dependent
on time because the system is quite far from equilibrium; however at $\tau \sim 0.8 \rm \, fm/c$ 
both the value and the time evolution become indistinguishable from the thermal cases 
represented by the Th-Glauber and Th-fKLN. 
We notice a little adjustment
also for these cases that we have indicated as thermal. The reason is that the initial spectra are
thermal only in the transverse plane, but there is a boost invariance in the longitudinal direction. This causes a little re-adjustment
that would disappear assuming a thermal spectral also in the longitudinal direction.

Our results on thermalization time are not in disagreement with earlier studies showing that two-body collisions are insufficient
to achieve a fast thermal equilibrium \cite{El:2007vg}. In fact in that case a perturbative QCD two-body cross section 
is employed, corresponding to $\eta/s$ about one order of magnitude larger than in our case \cite{Plumari:2012ep}, 
while here we normalize the cross section to get an $\eta/s=0.08$ that implies scattering rates 
very large that induce a fast thermalization.

\begin{figure}
\begin{center}
\includegraphics[width=7cm]{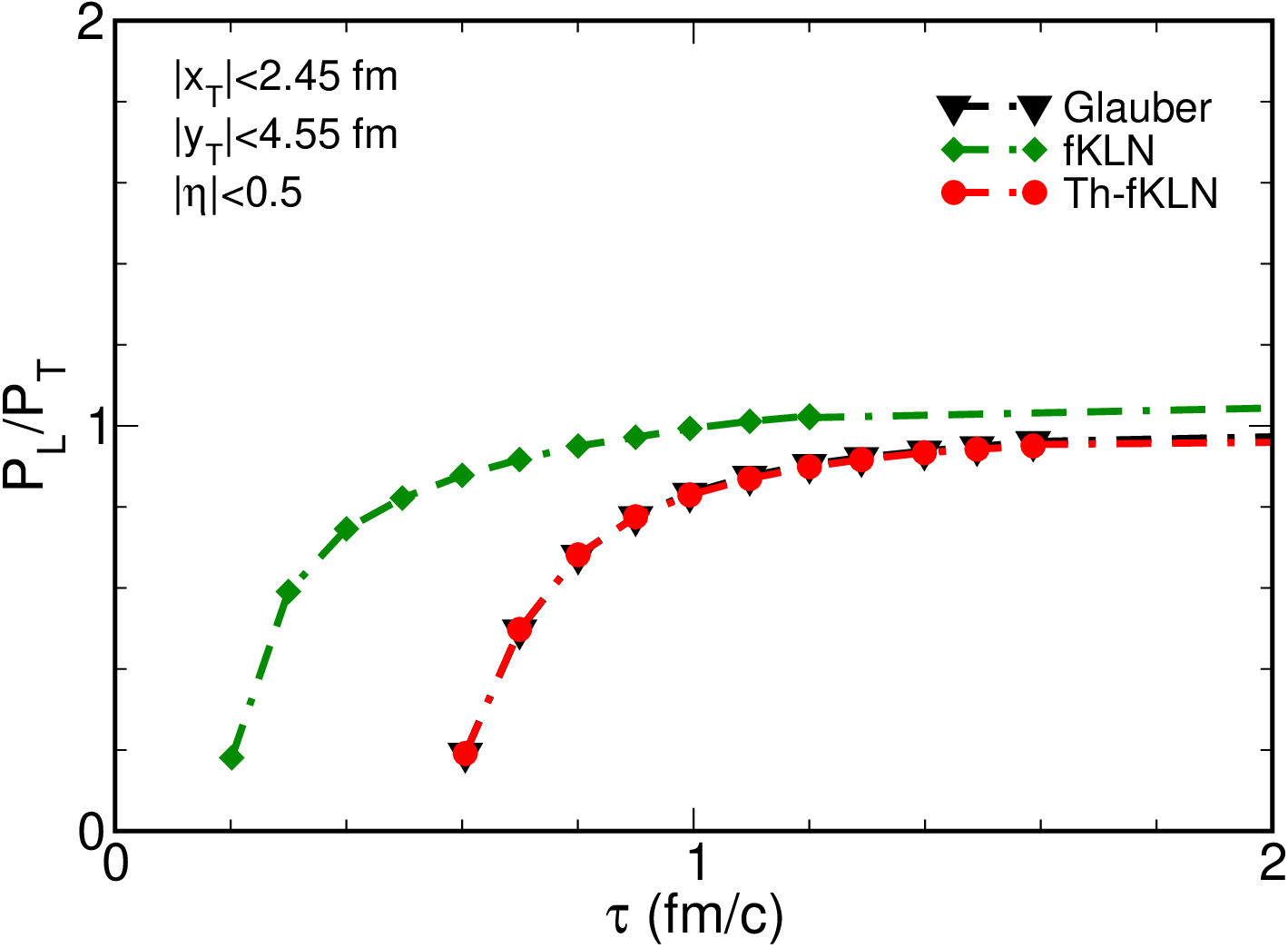}
\caption{\label{Fig:press}
Time evolution of the longitudinal pressure over the transverse pressure,
computed for the three different initial conditions discussed in the article.
Data correspond to Au-Au collisions
at $\sqrt{s} = 200$ GeV, with an impact parameter $b=7.5$ fm
and $4\pi\eta/s=1$.}
\end{center}
\end{figure}

We have studied also the isotropization of the pressures of the expanding system. 
In order to do this we have computed the relevant components of the energy-momentum tensor, 
\begin{equation}
T^{\mu\nu}=\int\frac{d^3 \bm p}{(2\pi)^3}\frac{p^\mu p^\nu}{E}f(x,p)~,
\label{eq:Tmunu1}
\end{equation}
where $f$ corresponds to the invariant distribution function. In our simulations the energy-momentum tensor is defined
in each cell; we then define transverse and longitudinal pressures, $P_T$ and $P_L$ respectively, as
\begin{eqnarray}
P_T&=&\frac{1}{V}\int_\Omega d^2\bm x_\perp d\eta~\frac{T_{xx} + T_{yy}}{2}~,~~~ \\
P_L&=&\frac{1}{V}\int_\Omega d^2\bm x_\perp d\eta~T_{zz}~,
\end{eqnarray}

where the integration is restricted to the region $\Omega$ defined by $|x|<2.45$ fm, $|y|< 4.55$ fm and $|\eta|< 0.5$,
and $V$ is the volume of such a region. In the lower panel of Fig.~\ref{Fig:press} 
we plot our results about $P_L/P_T$ as a function
of time, for the three initial conditions considered till now. As we have already explained, the initial time for the simulation
depends on the particular initial condition used. Our findings suggest that independently on the initial condition
implemented, the system becomes isotropic within 1 fm/c. A similar result has been obtained recently using both $2\leftrightarrow2$
and $2 \leftrightarrow 3$ collisions with pQCD cross sections and $\alpha_s=0.6$ which should correspond to $\eta/s \approx 0.1$ 
\cite{Zhang:2012vi}.
This also mean that the main effect we observe on the elliptic flow
is not driven by the isotropization itself.
This result is not in disagreement with other studies ~\cite{Gelis:2013rba,Fukushima:2013dma}
which show how the expanding glasma becomes almost isotropic within few fm/c if
the coupling is strong enough, or in general the dynamics sufficiently nondissipative~\cite{Ryblewski:2013eja}.

\begin{figure}
\begin{center}
\includegraphics[width=8.5cm]{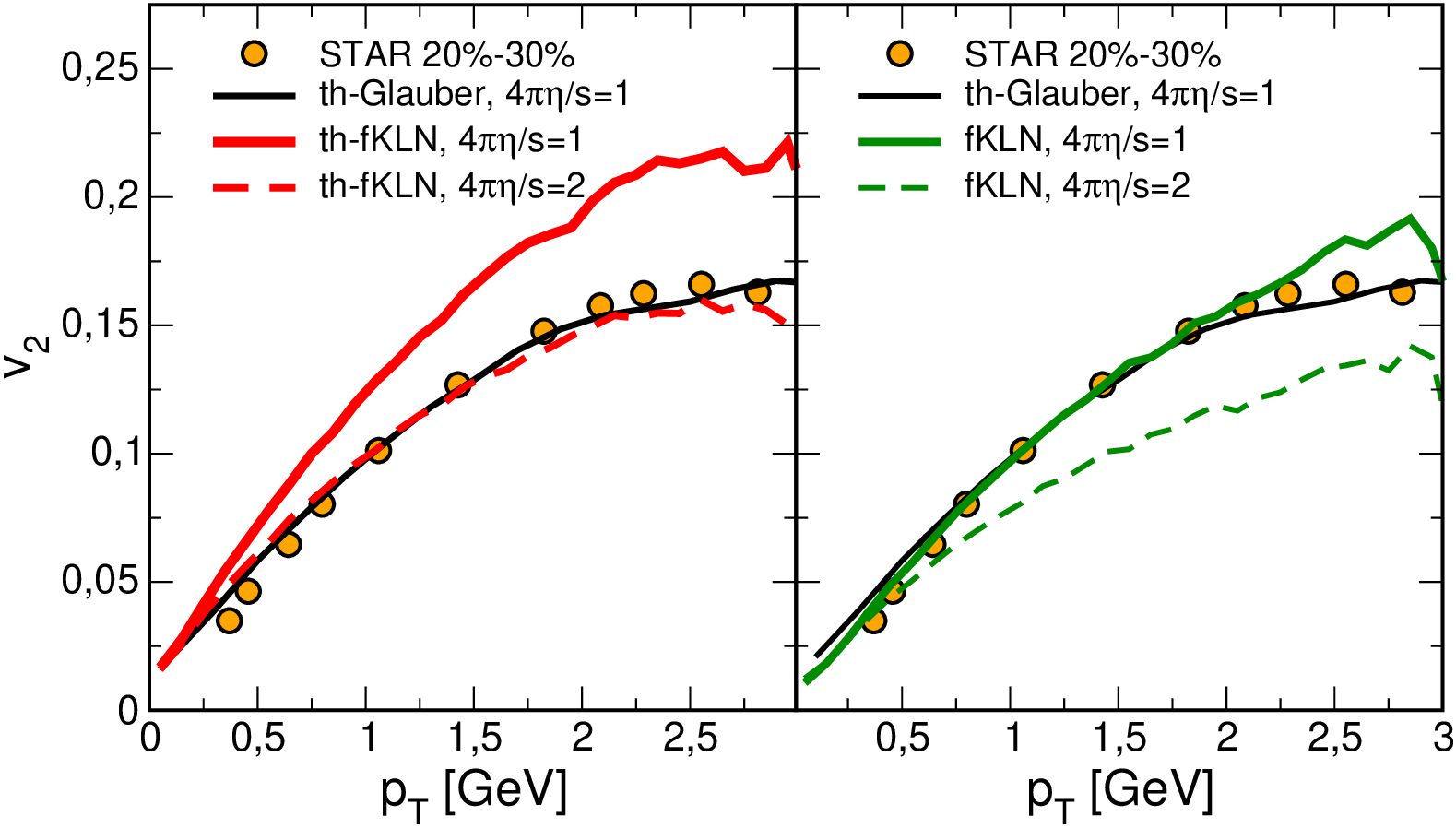}
\caption{ Elliptic flow $v_2(p_T)$ for the different initial conditions and $\eta/s$ as in the legend.
All the calculations refer to Au-Au collisions
at $\sqrt{s} = 200$ GeV, with an impact parameter $b=7.5$ fm.}
\label{Fig:v2pt}
\end{center}
\end{figure}

In the left panel of Fig. \ref{Fig:v2pt} we plot the $v_2(p_T)$ 
for the case of Th-Glauber (thin solid  line) and Th-fKLN (thick solid line) at a fixed $4\pi\eta/s=1$. 
The Glauber initial condition reproduces quite well the data (circles); 
in the case of Th-fKLN (thick solid line) one gets a too large $v_2$ and for such initial conditions the agreement 
with the data is achieved only if the $\eta/s$ is increased by a factor of two (dashed line).  
These results are in agreement with the ones obtained from viscous hydrodynamics 
\cite{Luzum:2008cw,Alver:2010dn,Song:2011hk,Adare:2011tg}, showing the solidity and
consistency of our transport approach at fixed $\eta/s$. 

In the right panel of Fig.~\ref{Fig:v2pt} we present our
novel result for the fKLN model, when the CGC distribution function is implemented 
in both the $\bm x$ and $\bm p$ spaces. 
We find that fKLN with a $4\pi\eta/s=1$ (thick solid line) gives a $v_2(p_T)$ quite similar
to the Th-Glauber, while in such a case if $4\pi\eta/s=2$, dashed line, the $v_2(p_T)$ would be too small.
Our interpretation is that the initial larger $\epsilon_x$ is compensated by the key feature of an almost
saturated initial distribution in $\bm p$-space below the saturation scale $Q_s$.
In other words the initial out-of-equilibrium fKLN distribution reduces the efficiency
in converting $\epsilon_x$ into $v_2$. 
In fact, the elliptic flow can be understood as a larger slope of the momentum spectrum in the 
out of plane $\vec x$ direction with respect to the $\vec y$ one caused by a larger pressure in the $\vec x$ 
direction due the elliptical shape.
The net effect in terms of the difference of the particle yields between the two directions is larger if the spectra
are decreasing exponentially with respect to the case in which they are nearly flat as a function of $p_T$.
A detailed study is in preparation, but the result we present in this Letter shows that 
the initial out-of-equilibrium function implied by the CGC cannot be discarded in studying the build of the
$v_2$ and the effect change significantly the estimate of  $\eta/s$.

We however notice that in the context of viscous hydrodynamics it is possible to include initial non-equilibrium conditions by introducing an initial non-vanishing value for the viscous tensor $\Pi^{\mu \nu}(t_0)$. This has been seen to have a quite small impact on the $v_2$ at least on the bulk of the system \cite{Luzum:2008cw,Song:2007fn,Niemi:2011ix}. However to our knowledge it has never been investigated the relation, if any, between the non-equilibrium implied by $\Pi^{\mu \nu}(t_0)$ and the one implied by the saturation scale in the distribution function $f(x,p)$ in CGC inspired models.

\begin{figure}
\begin{center}
\includegraphics[width=7.3 cm]{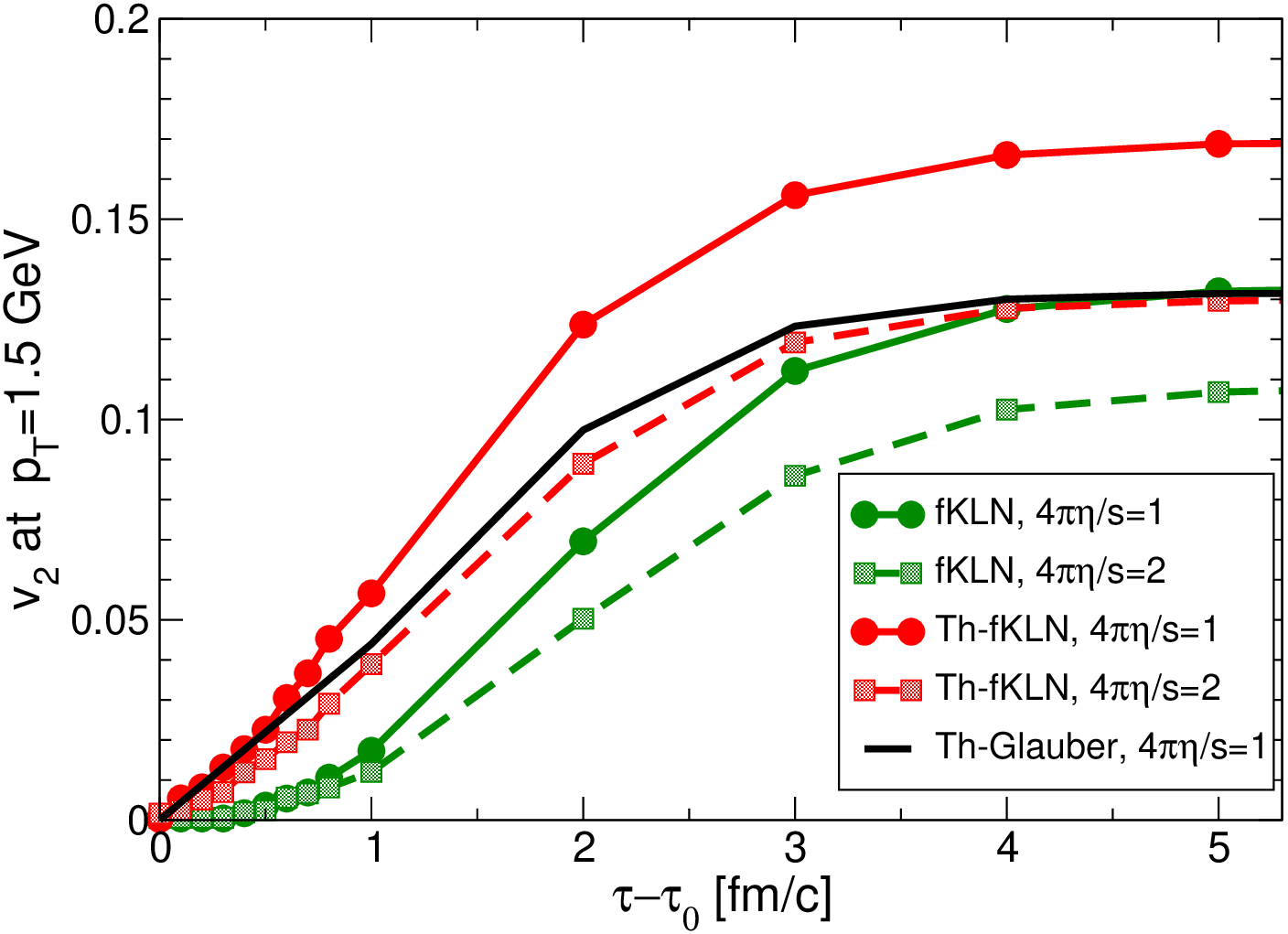}
\caption{ Evolution of $v_2$ at $p_T=1.5 \rm \, GeV$ as a function of the evolution time 
for all the different initial conditions and $\eta/s$ values.
 Calculations refer to Au-Au collisions
at $\sqrt{s} = 200$ GeV, with an impact parameter $b=7.5$ fm.}
\label{Fig:v2_time} 
\end{center}
\end{figure}

In Fig. \ref{Fig:v2_time}, we show the build-up of $v_2$ at a fixed $p_T=1.5 \rm \, GeV$  as a function
of time, $\tau-\tau_0$, for all the initial conditions considered. 
We find that the rate of increase of $v_2$ 
for the thermal distributions is large from the very beginning of the evolution, 
while it is quite reduced for fKLN. After about $1-1.5 \rm fm/c$, 
roughly corresponding to the thermalization also for the full fKLN distribution,
the increase of $v_2$  with time also for fKLN becomes very similar to Th-fKLN. 
We notice that when plotted as a function of $\tau-\tau_0$ 
the evolution of the space eccentricity is quite similar between $Th-fKLN$ and $fKLN$,
hence the differences observed are mainly driven by the p-space distributions.
This observation further confirms that it is the initial
out-of -equilibrium and nearly saturated distribution that dampens the efficiency in converting the 
space eccentricity into the $v_2$. Therefore, even if the thermalization sets in quite quickly,
as assumed in hydrodynamics $\tau \sim 0.8 \rm \, fm/c$, such a time duration cannot be neglected 
in studying the impact of initial conditions.

In conclusion, our study based on kinetic theory shows that the elliptic flow in a system
depends not only on the pressure gradients and the $\eta/s$ of the system, but also on the
initial distribution in momentum space. 
A $\bm p$-distribution with a saturation behavior generates smaller $v_2$ with respect to the thermal one.
This result is quite general, and we expect it should be valid, besides QGP in uRHICs, 
for systems like cold atoms in a magnetic trap which are characterized by a value of $\eta/s$ 
close to the QGP one \cite{O'Hara:2002zz}.
In the specific case of the KLN matter studied here, 
the effect of the initial non-equilibrium distribution affects the estimate of $\eta/s$ 
of about a factor of two.
The relevance of our results is further enhanced by the fact that Th-fKLN with 
$4\pi\eta/s \sim 2$ would generate a low $v_3$ with respect to the available data, which is
the main conclusion of~\cite{Adare:2011tg}.
We notice that in the present kinetic approach the quantum nature of gluons
has been discarded. This is justified at RHIC because 
in this Letter we have focused on an effect which is dominant at $p_T> 0.5$ GeV,
where the $f(x,p)$ is still smaller than unity. At LHC, or anyway at small $p_T$,
 it would be necessary to
include $(1+f)$ terms in Eq.(\ref{eq:BE}) that could drive the system toward a Bose-Einstein condensate
\cite{Blaizot:2013lga}.  

{\em Acknowledgements.}
The authors acknowledge L. Albacete and T. Hirano for useful suggestions and 
correspondence and D. Kharzeev for enlightening discussions.

\end{document}